\documentclass[aps,prl,preprint,tightenlines]{revtex4}
\begin{document}

\title{All Entanglements in  a Multipartite System}

\author{Yu Shi}

\email[Email:]{ys219@phy.cam.ac.uk}

\affiliation{
Cavendish Laboratory, University of Cambridge,
Cambridge CB3 0HE, United Kingdom} 

\affiliation{Department of Applied Mathematics and
Theoretical Physics, University of Cambridge, Cambridge CB3 0WA, 
United Kingdom
}

\begin{abstract}

For a multipartite system, 
we sort out all possible entanglements, each of which is 
among a set of subsystems. 
Each entanglement can be measured by a 
generalized relative entropy of
entanglement, which is  conserved on average 
under reversible local operations and classical
communication (LOCC) defined for all the parties.  
Then  we derive a series of inequalities 
of different entanglements that have to be satisfied by
any pure state which  can be generated by  reversible 
LOCC from the set of all GHZ-like states. 


{\bf PACS}: 03.67.-a, 03.65.Ud
\end{abstract}
  
\maketitle

\section*{1. Introduction} 

Entanglement is an essential quantum feature which lies at the heart 
of quantum mechanics~\cite{einstein} and is also a crucial 
resource of quantum information processing~\cite{bd}.  
Many recent  results in characterizing entanglement 
follow the idea that  entanglement is 
not  changed by any local unitary transformation,
and on average cannot be increased by local operations and classical
communication (LOCC)~\cite{bennett1,bennett2,vedro},
hereby called LOCC non-increasing principle.
In the bi-partite case, 
the entanglement is just  between the two parties, though there may be
different measures for a mixed state.
It is known that 
$p$ copies of a bi-partite pure state $\psi$, whose von Neumann entropy of 
either  party  is $E(\psi)$, 
can be  transformed by reversible LOCC
into $pE(\psi)$ copies of EPR singlet
states asymptotically, i.e.
in the limit of large $p$~\cite{bennett1}.  
Thus $E(\psi)$ 
is the measure of the bi-partite entanglement in the pure state
$\psi$.  One can use the term 
{\em LOCC equivalence} to refer  to the transformation  
under reversible LOCC, with certain number of copies for each
state involved in the transformation. 
Thus  for a bi-partite system, any entangled pure
state is LOCC asymptotically  equivalent to EPR state. 

For a  multipartite system, how to characterize 
the  nature of entanglement has remained  quite
open, with many interesting questions, for example,  how many
different kinds of entanglement there exist and how different
entangled pure states  
are transformed into each other by reversible LOCC. 
First of all,
what does a $N$-partite entangled state precisely mean?  

Let us refer to the state  
$$g(1\cdots n) \equiv \frac{1}{\sqrt{2}}(|0_1\rangle|\cdots|0_n\rangle+
|1_1\rangle|\cdots|1_n\rangle)$$
  as {\em $n$-GHZ state} or  
{\em a GHZ-like state},
where the subscripts represent the different parties. Hence 
$2$-GHZ state is just the EPR state. 
Recently it was shown that
$4$-GHZ state  is not LOCC  
asymptotically equivalent to any combination of the
six different EPR pairs and that 
two different  $3$-GHZ states is not exactly LOCC equivalent 
to   three EPR pairs~\cite{bennett3}. It was
also  shown that 
$N$-GHZ is not LOCC asymptotically equivalent  
to any combination  of $k$-GHZ states, for all  $k<N$~\cite{linden}.

An interesting concept is the {\em reversible  
LOCC entanglement  generating set} (RLEGS)~\cite{bennett3}. 
For  a given multipartite system, a set of different states 
is a RLEGS if a  combination of  these states, 
each entering  with certain  copies,  
can be transformed, under reversible LOCC, into  
certain copies of an  arbitrary entangled pure state. The 
numbers of copies of the states concerned could be arbitrarily large. 
For example, EPR state comprises a RLEGS for a bi-partite system. 
An interesting  question is that for a system of 
$N>2$ parties,  whether 
all kinds of GHZ-like states, each shared by a
subset of the parties, is a RLEGS.  
A negative answer is known for $N=4$~\cite{wu}.

For a given   $N$-partite system, let us define 
a  proper subset of parties as a  {\em generalized parties} (GP).
A ``proper subset'' means that the number of
elements in the subset is less than $N$. 
The Hilbert space of each GP is the tensor
product of those of the parties belonging to it. 
The notion of LOCC can be  generalized to the  generalized LOCC
(GLOCC), which means operations that are local
with respect to the GPs and communication among the GPs. 
We  shall sort out all  different 
entanglements in a  $N$-partite system, 
each corresponding to  a  different set of at least two GPs.
Thus 
{\em a  $N$-partite entangled state is a $N$-partite state 
in which  at least one of the entanglements is nonzero}. 
The LOCC-nonincreasing principle appropriately formulated for
a $N$-partite system   leads to the 
conservation of any  {\em entanglement  involving
all the $N$ parties} (see below for precise
meaning),  under reversible LOCC. 
This  approach  reproduces the conclusion 
about the existence of the so-called  true 
$N$-partite entanglement, with which
a state cannot be transformed,
by using reversible LOCC, 
into states in which only $k<N$ parties are entangled \cite{linden}. 
Moreover,  it is identified as  the   
{\em entanglement  among  all the $N$  parties}  (see below for precise
meaning).   For each entanglement,  we introduce a 
{\em generalized relative entropy} (GRE)   as a 
measure.  GRE of each entanglement is on average
conserved under reversible  LOCC. 
Based on these conservation laws, we consider an arbitrary 
$N$-partite pure state, with certain copies,
which can be generated, under reversible LOCC, 
from the set of all GHZ-like states.  We obtain a series of
inequalities of different entanglements, which must be
satisfied by any such  state. 

\section*{2. Entanglement among a set of generalized parties}

Entanglement is a kind of correlation,
among at least  two objects. Thus 
we can consider all
possible entanglements  in a $N$-partite system,   
by {\em associating an  entanglement with 
each  set of at least two GPs}, called the 
entanglement among, or of,  a set of GPs. 
When we say ``a set of GPs'', we always mean a set of nonoverlapping
GPs, i.e.   there is no party shared by different GPs. 
We use a number, sometimes put in a
bracket,  to represent each party,  a
sequence of numbers in a bracket to represent the GP consisting of
these  parties, and  a succession of brackets as a set of GPs. 
For example, 
for a system of  three parties  $(1)$, $(2)$ and $(3)$,  there are
entanglements of $(1)(2)$, $(1)(3)$, $(2)(3)$,
$(1)(23)$, $(2)(13)$, $(3)(12)$ {\em  and $(1)(2)(3)$}, respectively. 
For a $N$-partite system, 
the total number of different sets  of 
at least two GPs is 
\begin{equation}
\sum_{n=2}^N \frac{N!}{n!(N-n)!}\sum_{m=2}^n P(n,m),
\end{equation}
where $P(n,m)=\sum_{n_1,\cdots,n_m}\frac{n!}{n_1!\cdots n_m! m!}$,
with  $n_1+\cdots+n_m=n$, is the number of different partitions of
$n$ parties into $m \geq 2$ GPs,  denoted as 
$(1^1\cdots 1^{n_1})(2^1\cdots 2^{n_2})\cdots(m^1\cdots m^{n_m})$, 
where $q^{i}$ denotes the $i$-th party  in the $q$-th GP.  
Note that in this terminology, the amount of 
an entanglement could  be zero. 

Now we introduce  {\em  generalized local operation}  
and {\em generalized LOCC} (GLOCC). 
A generalized local operation  is an operation on  
a GP. It is local with respect to the GP, but may
be nonlocal with respect to the parties. 
One knows that a process of LOCC transforms the state as  
$\rho \rightarrow \sum_{k}  L_k \rho
L_k^{\dagger}$, where $L_k=\otimes_{i=1}^N l_k^i$, 
$l_k^i$ is a  local operation  on party  $i$, with 
$\sum_k {l_k^i}^{\dagger}l_k^i \leq 1 $.
Similarly, a process of 
GLOCC, 
on  $m$ GPs  corresponding to a  partition of $N$ parties, 
 transform the state as 
$\rho \rightarrow \sum_j  G_j\rho
G_j^{\dagger}$, where $G_j=\otimes_{q=1}^m g_j^q$, 
$g_j^q$ is a generalized local operation  on  GP  $q$,
with  $\sum_j {g_j^q}^{\dagger}g_j^q \leq 1 $.  

A local operation  on a party
is also local with respect to a any GP to which the party
belongs, while  communication among all the parties is also 
communication among all the GPs corresponding 
any partition of the parties. 
Therefore, a process of  LOCC  is also a process of 
GLOCC corresponding to any partition of the parties to GPs.
 $L_k=\otimes_{i=1}^N l_k^i$ can also  be written as
a product of generalized local operations on different 
GPs under {\em any} partition, i.e. 
$L_k=\otimes_{q=1}^m g_k^q$, where  $g_k^q =  \otimes_{i\in
q}l_k^i$. 
However, 
a generalized local operation on a GP may not be a product of 
local operations on parties belonging to this GP.

A LOCC process  for  $N$ parties may not be a LOCC process
on a proper subset of the parties, i.e. $n$ of the $N$ parties, $n <
N$, while a LOCC process  for a proper  subset  of $N$ 
parties is still a LOCC  process for the  $N$ parties. 
If one treat the system as two parts $A$ and $B$, then 
an operation $L_k$ can be written 
as $L_k^A \otimes L_k^B$, where 
$L_k^A=\otimes_{i\in A} l_k^i$,  $L_k^B=\otimes_{n\in B} l_k^i$.
Under a process of LOCC on the total system, 
the reduced density matrix of $A$ evolves as
$\rho_A \rightarrow tr_B( \sum_k L_k\rho L_k^{\dagger})
=\sum_k L_k^{A}tr_B(L_k^B \rho L_k^{B^{\dagger}}) L_k^{A^{\dagger}}
\leq \sum_k  L_k^{A} \rho_A L_k^{A^{\dagger}}$. 
Similar is the  situation of GLOCC. 

It is straightforward to obtain the following.  
(i) A generalized local operation
on a GP does not change  the reduced density matrix
of any other nonoverlapping GP.
(ii) GLOCC among a set of  GPs  do not change  the reduced
density matrix of any other  GP nonoverlapping with them. 
(iii) The von Neumann entropy of a GP
is  non-increasing under GLOCC 
for any set of GPs including the concerned one. 

There may  be various 
relations among different entanglements.
After all,  each entanglement is 
a function of the same  state. 
Consider one copy of a $N$-qubit pure state, which can be  specified by 
$2^{N+1}-2$  real   parameters,
up to a global phase. 
For sufficiently large $N$, the number of entanglements
is  larger than the number of parameters, thus 
there must  be relations among different entanglements.
The relations  depend on the state.   
However, there are some qualitative
relations independent of the state, as reflected in the features
of the set of  the separable states in defining the  
GRE. For example, if a subsystem of a
GP is entangled with a subsystem of another GP, these two GPs must be 
entangled.  One can observe that it is impossible to have a state
for which only one of the entanglements   is
nonvanishing,  otherwise it would be straightforward to obtain a
RLEGS.

\section*{3. Entanglements involving all $n$ parties
and entanglement among all $n$ parties}

For $n$ parties $(1)$, $(2)$, $\cdots$, $(n)$,  we refer to 
the entanglement of $(1)(2)\cdots (n)$ as the {\em entanglement among all 
the $n$ parties}. In other words, for this entanglement,  
each party is a GP in the definition of the entanglement. 
 
If  the set of GPs,   in the definition of an entanglement, 
ultimately consists of all $n$  parties,  we refer to  the entanglement as 
``{\em an entanglement involving all  the $n$ parties}''.
Obviously, the first example is just the entanglement among all the
$n$ parties. Other examples include
 $(12)(3\cdots n)$,  $(12)(345)(6\cdots n)$,
etc.

In a  $n$-partite Schmidt decomposable 
pure  state 
$\sum_k a_k|\eta_k(1)\rangle|\cdots |\eta_k(n)\rangle$, where
$|\eta_k(i)\rangle$ represents the $k$-th basis state of party $i$,
the only nonzero entanglements are those  
involving all the $n$ parties.
Tracing out any number of parties leads to a completely
separable state.   A  $n$-GHZ state 
is an example  of $n$-partite Schmidt decomposable  state.  

\section*{4. LOCC  non-increasing principle and its consequences}

The LOCC non-increasing principle for a $N$-partite system,
as the straightforward generalization of $N=2$ case, 
can be formulated as the following. 

{\em LOCC non-increasing principle}----For a $N$-partite system,   
on average  the entanglement  among  all the 
$N$ parties  is  invariant under local unitary
transformations  and  cannot be increased by  
LOCC  for  {\em all} these $N$ parties.

By  {\em ``coarse graining''}, i.e. 
replacing the parties as GPs while LOCC as GLOCC,
the above principle also implies the following. 
(i) A generalized local unitary transformation 
on a GP  does not change 
any  entanglement  among  a set of GPs one of which 
is this  GP,  and does not change any
entanglement  among any other GP nonoverlapping with this GP.
(ii) GLOCC among  a set of GP
does not increase   the entanglement  among all  these GPs. 

These results, together with 
the fact that a process of 
LOCC is also a process of GLOCC with respect to any partition,
lead to the conclusion that
LOCC on a multipartite system 
does not increase {\em any} entanglement 
{\em involving all} the parties.

This deduction is like that of renormalization
group: making coarse graining and
then finding the fixed points. 
Here  coarse graining is  regarding a set of parties as 
a basic unit, i.e.  partitioning a set of  parties into a GP. In doing so, 
a process of GLOCC, previously corresponding to a certain partition,
may not be a GLOCC process
 corresponding to the  coarse graining made.
However, any process of LOCC is still a
valid process of GLOCC corresponding to the coarse graining made.  
Hence a process of LOCC is a fixed point under any  coarse graining. 

Therefore we have obtained the following conclusion. 

{\bf Theorem 1:} Each  entanglement 
involving all the parties is on average  conserved in a 
reversible process of LOCC for {\em all} these  parties.
 
Consider  two  pure states of $N$ parties,
which are partitioned into $m$   GPs.   By 
generalizing a theorem in \cite{bennett3}, one knows that 
if for the two pure states, 
each GP has a  same von Neumann entropy,
then  these two states are either GLOCC incomparable or 
equivalent under the generalized local unitary transformations,
with respect to the given partition. 
From this, we obtain the following.

{\bf Theorem~2:} If two $N$-partite pure 
states are LOCC equivalent, then
they are  equivalent under the  generalized local unitary transformations
with respect to any  partition.  

{\bf Proof:} 
As a special case of Theorem~1,
LOCC equivalence implies that for these two states,
any GP has a same entropy, which measures the bi-partite entanglement 
between this GP and the rest  parties. 
Thus  these two states are either 
equivalent under the generalized local unitary transformations, 
or GLOCC incomparable, with respect to any partition.
The latter possibility can be excluded
by the fact that a  LOCC process
is a also process of  GLOCC corresponding to any partition of 
parties into GPs. 

The so-called
{\em true $n$-partite entanglement} can be
identified  as
nothing but  the entanglement among all the $n$ parties, in the
following way.
First the  true $n$-partite
entanglement
must involve all the $n$ parties. Then
consider an entanglement 
involving all the $n$ 
parties but is not the one among all the $n$ parties. 
{\em Such an entanglement can be contributed by a state 
shared by less  than $n$ parties, if the state is shared among all the
GPs}. 
For example, the entanglement between GPs $(12)$ and
$(3\cdots n)$ is contributed by 
a bi-partite entangled state shared between parties $(2)$ and $(3)$.
Indeed,  a process of LOCC for  $(2)$ and $(3)$ 
is also a process of GLOCC for $(12)$ and
$(3\cdots n)$.  On the other hand, when each of the $n$ parties  is
a GP, GLOCC is just LOCC.   Therefore among all the  
 entanglements involving all the $n$ parties, only 
the entanglement among all the $n$ parties is the true $n$-partite
 entanglement, according to its definition. 
Theorem~1 guarantees the LOCC inequivalence of different entanglements
involving all the parties (each of them is conserved {\em
respectively} under reversible LOCC defined for these $n$ 
parties, hence there is not  
a situation that one of them is transformed into another). 
A corollary  is that 
$N$-GHZ state, with nonzero true $N$-partite entanglement, 
is not LOCC equivalent to any
repertoire of $k$-GHZ states for all $k<N$. 

Note that for a $N$-partite system, there 
exists a true $n$-partite entanglement for any 
$n$ parties,  $2\leq n \leq N$.

\section*{4. Generalized relative entropy of entanglement}

For a $N$-partite system, consider $n\leq N$
parties, with the  (reduced)  density matrix $\rho$. 
The generalized relative entropy of entanglement (GRE) defined for   
the entanglement among $m$ GPs $(1^1\cdots 1^{n_1})$,
 $(2^{1} \cdots 2^{n_2})$, $\cdots$, and $(m^1,\cdots,m^{n_m})$, with
$n_1+\cdots+n_m =n$,  is
$$E'(\rho)= Min_{\sigma} Tr(\rho \ln \rho-\rho\ln \sigma),$$ 
where  the separable density matrix
$\sigma$ is  of the form
\begin{equation}
\sigma =\sum_s \sum_{k_s} p^s_{k_s} \otimes_{\alpha_s} 
\sigma^{(\alpha_s)}_{k_s}  \label{sep}
\end{equation}
where $s$ denotes each different  partition of the $m$ GPs 
{\em further  into various sets of GPs} numbered as $\alpha_s$
(each partition $s$ represents a way of separation), 
$k_s$ denotes  the decomposition of
density matrix under the partition $s$,
$\sigma^{(\alpha_s)}_{k_s}$ is  in the
Hilbert space of $\alpha_s$ and can be required to be pure.
 {\em For a given} 
$\sigma$,   $\sigma$ is separable 
with respect to a certain partition of
GPs into sets of GPs (of course, some separable states
can be separated in more than one way), i.e.
  $p^s_{k_s}\neq 0$ only for  $s=s_0$. 
Hence   $\sum_{k_{s_0}}p^{s_0}_{k_{s_0}}=1$ while 
$p^{s}_{k_{s}}=0$ for $s\neq s_0$.
In other words, {\em the set of separable states is the union of all
sets of  the states 
separable with respect to different partitions of the $m$
GPs}. 
The reason to  consider the separability of all different ways
is to exclude the entanglements among not
all the  $m$ GPs.

Take  a density matrix  of three parties $1$, $2$ and $3$ 
as an illustration. 
For  the GRE of
 $(1)(2)(3)$,
$\sigma=
\sum_{k_1} p^1_{k_1}\sigma^{(1)}_{k_1} \otimes \sigma^{(2)}_{k_1}
 \otimes \sigma^{(3)}_{i_1}$ 
{\em or}  
  $\sigma= \sum_{k_2} p^2_{k_2} \sigma^{(12)}_{k_2}  
\otimes \sigma^{(3)}_{k_2}$ {\em or}  
 $\sigma= \sum_{k_3} p^3_{k_3} \sigma^{(13)}_{k_3}  \otimes 
\sigma^{(2)}_{k_3} $  {\em or}  
 $\sigma= 
\sum_{k_4} p^4_{k_4} \sigma^{(23)}_{k_4}  \otimes \sigma^{(1)}_{k_4}$.
For   the GRE  of 
$(12)(3)$, 
$\sigma = \sum_k p_k \sigma^{(12)}_k  \otimes \sigma^{(3)}_k$.
For  the GRE of $(1)(2)$, 
$\sigma= \sum_k p_k \sigma^{(1)}_k  \otimes \sigma^{(2)}_k$.

For convenience, let us 
normalize the  GRE 
of  $n$-parties as 
$$E(\rho)=E'(\rho)/E'_0,$$
 where $E'_0$ is  for $n$-GHZ
state, such that $E=1$ for  $n$-GHZ
state.  Clearly the GRE satisfies the
LOCC non-increasing principle. 

The GRE  of {\em each}  entanglement in a $N$-partite {\em pure} state  
is on average  conserved under reversible
LOCC process.
For an 
entanglement involving all the parties, this has been stated in
Theorem~1.
An  entanglement not involving all the parties 
{\em  can} 
be increased  under LOCC  for all the $N$ parties, 
but its GRE 
is still  conserved on average under reversible
LOCC, as a result of
\cite{linden},  seen as follows.
Consider  the GRE for an entanglement  involving $n$ parties
out of the $N$ parties,  $2 \leq n <N$. 
Any one of the $n$ parties can act as the Bob in
\cite{linden}, while any one of the other $N-n$ parties can act as the
Alice.  Then a straightforward use of 
the proof in \cite{linden} shows that 
the average increase of the GRE of any entanglement 
involving the $n$ parties 
is not larger than the average decrease of the von Neumann entropy 
of the GP  consisting of these $n$ parties.  

Obviously, subadditivity and monotonicity~\cite{werner}
holds for GRE, since they do
not  depend on the details of the separable density matrices in the
definition of the relative entropy of entanglement. Subadditivity
means $E(\rho_a\otimes\rho_b)\leq E(\rho_a)+E(\rho_b)$
for any $\rho_a$ and $\rho_b$.
Monotonicity means $E(\rho_a\otimes\rho_b)\geq E(\rho_a)$ if
$E(\rho_b)=0$. The combination of these two properties means
$E(\rho_a\otimes\rho_b)= E(\rho_a)+E(\rho_b)$ if $E(\rho_b)=0$~\cite{werner}.
Besides, for an entanglement between
a GP and its complementary GP, i.e. the rest of system, 
 the GRE reduces to a von Neumann entropy and is thus additive. 

\section*{5. Constraints on any state which can be reversibly 
generated by LOCC from the set of all GHZ-like states} 

A GHZ-like state is a sort of ``canonical'' entangled state, hence it
is interesting to consider the set of all  GHZ-like states
for $N$ parties, denoted as  ${\cal G}(N)$.
Each of these GHZ-like states is 
shared by a different set of parties.
For $N$ parties, 
there are altogether $M_{ghz}=2^N-N-1$ GHZ-like states.

Suppose a $N$-partite state can be reversibly 
generated by LOCC from ${\cal G}(N)$. This means 
that 
$c_{\psi}$ copies of $|\psi\rangle$ can be converted, 
by using  reversible LOCC, into a set of  
$c_z$ copies of  state  $|z\rangle$, with each $|z\rangle$
belonging to  ${\cal G}(N)$. 
$c_{\psi}\neq 0$, $c_z$ could be arbitrary non-negative integers,
and not all $c_z$-s are zero.  
In the following, we shall give some necessary conditions for
$|\psi\rangle$.
If an arbitrary 
$N$-partite entangled pure state can be  reversibly 
generated by LOCC from  ${\cal G}(N)$, then  ${\cal G}(N)$ is 
a RLEGS. Hence if any of the necessary conditions about $|\psi\rangle$
can be violated by any $N$-partite entangled state, then ${\cal G}(N)$ 
is not a RLEGS. 

The conservation of GRE for each entanglement implies 
\begin{equation}
E_{\alpha}( \otimes_z |z\rangle^{\otimes c_z}) 
= E_{\alpha}(|\psi\rangle^{\otimes c_{\psi}}), \label{e}
\end{equation}
where $\alpha$ denotes the entanglement corresponding to
each different set of GPs,  $E_{\alpha}(\phi)$ represents the 
corresponding GRE possessed by state $\phi$.  
Subadditivity of GRE implies 
\begin{equation}
E_{\alpha}( \otimes_z |z\rangle^{\otimes c_z})
\leq \sum_{z} c_z E_{\alpha}(|z\rangle),  \label{z} 
\end{equation}
and
\begin{equation}
E_{\alpha}(|\psi\rangle^{\otimes c_{\psi}})
\leq c_{\psi} E_{\alpha}(|\psi\rangle), \label{ps} 
\end{equation}
for any entanglement $\alpha$. 

It has recently been known that in general, 
relative entropy  is not 
additive~\cite{werner}. Thus (\ref{e}) cannot be reduced a systems of 
equations of $\{c_z\}$ and $c_{\psi}$, which would be easy to handle. 
However,   some of the entanglements are additive,
leading to some simplification.

First we note that for the combination of GHZ-like states,  
$\otimes_z |z\rangle^{\otimes c_z}$, 
an  entanglement is only 
contributed by those GHZ-like states only involving  
a subset of each of the  GPs in defining the entanglement. 

Then we  consider the  entanglement between
two  GPs,  referred to as a bi-GP entanglement,
in $\otimes_z |z\rangle^{\otimes c_z}$.
{\em If a  bi-GP entanglement is  contributed by  
a certain GHZ-like state, it  must be from the entanglement between two 
complementary parts in this GHZ-like state, and thus
must be a von Neumann entropy, which is additive}. 
On the other hand, trivially 
if a GHZ state is only held by one of the two GPs, 
it does not contribute this bi-GP entanglement. 
For example, to calculate the amount of 
entanglement between party $(1)$ and $(2)$ in a GHZ-like  
state,  one first  traces out any party 
which is neither $(1)$ nor $(2)$, and then consider the reduced density
matrix shared by party $(1)$ and $(2)$. Thus 
the entanglement between party $(1)$ and $(2)$ equals                          $1$ in state $g(12)$, but is zero in any other GHZ-like state, 
e.g. $g(13)$, $g(23)$, $g(123)$, etc. As another example, the 
entanglement between GP $(12)$ and $(3)$ equals $1$ in $g(123)$, $g(23)$
and $g(13)$,  while
is zero in any other GHZ-like state, e.g. $g(12)$, $g(1234)$,
$g(234)$, etc.  

Therefore for any GHZ-like state, any bi-GP entanglement  either
vanishes or is a von Neumann entropy of the reduced density matrix of 
either of the two GPs. 
Consequently {\em  any 
bi-GP entanglement is additive for $\otimes_z |z\rangle^{\otimes
c_z}$}.

Hence for each bi-GP entanglement, LHS of Eq.~(\ref{e}) becomes
\begin{equation} 
E_{bG}( \otimes_z |z\rangle^{\otimes c_z}) 
= \sum_z c_z E_{bG}(|z\rangle), \label{z1}
\end{equation}
where the subscript $bG$ represents  bi-GP entanglements.  

Combining  Eq.~(\ref{z1}) and Eq.~(\ref{ps}), one obtains 
\begin{equation}
\sum_z x_z E_{bG}(|z\rangle) \leq  E_{bG}(|\psi\rangle), \label{ine}
\end{equation}
where $x_z \equiv c_z/c_{\psi}$. 
The total number of different
bi-GP entanglements is $M_{bG}=\sum_{n_1=1}^{N-1} 
\sum _{N_2=1}^{N-n_1} \frac{N!}{n_1!n_2!(N-n_1-n_2)!}$, which is  
larger than the total number of GHZ-like states $M_{ghz}$.

Moreover,  for pure
states, e.g. copies of  $|\psi\rangle$  on RHS of Eq.~(\ref{e}),
a  bi-GP entanglement   {\em involving all
the parties},  i.e. the entanglement  between a GP and its complementary GP, 
as a von Neumann entropy,  is additive.   
There are $M_{bGa}=2^{N-1}-1$ bi-GP entanglements
involving  all the parties. The subscript $bGa$ represents
bi-GP entanglements involving  all the parties.  Therefore one obtains 
$M_{bGa}$ equations, 
\begin{equation}
\sum_z x_z E_{bGa}(|z\rangle) =   E_{bGa}(|\psi\rangle), \label{bga}
\end{equation}
in addition to $M_{bG}-M_{bGa}$ inequalities given by  Eq.~(\ref{ine}). 

In terms of the GHZ-like states with nonzero amounts of the 
entanglement in consideration,   Eq.~(\ref{ine}) reduces to   
\begin{equation}
\sum x_{g(j_1\cdots j_l,j_{l+1}\cdots j_k)}
\leq  E_{(i_1\cdots i_r)(i_{r+1}\cdots i_n)}(|\psi\rangle) \label{bg1}
\end{equation}
where $2\leq n \leq N$,  $1 \leq r \leq n-1$,
 $(i_1\cdots i_r,i_{r+1}\cdots i_n)$ represents
any $n$ parties.
The summation is over all such GPs 
$(j_1\cdots j_l,j_{l+1} \cdots j_k)$  in  which
$(j_1 \cdots j_l)$ is a subset of GP $(i_1\cdots i_r)$ while 
$(j_{l+1}\cdots j_k)$ is a subset of GP $(i_{r+1}\cdots i_n)$. 
Eq.~(\ref{bga}) reduces to 
\begin{equation}
\sum x_{g(j_1\cdots j_l,j_{l+1}\cdots j_k)}
= E_{(i_1\cdots i_r)(i_{r+1}\cdots i_N)}(|\psi\rangle) \label{bga1}
\end{equation}
for any  $1 \leq r \leq N-1$.

Besides the bi-GP entanglements, which we have considered so far, it
is also useful to note that in $\otimes_z |z\rangle^{\otimes c_z}$,
the entanglement among all $n$ parties, 
$2 \leq n \leq N$, i.e. the entanglement of $(i_1)\cdots(i_n)$,
where each $i_j$ is a party,  which we have called a true $n$-partite 
entanglement,  is only contributed by
$g(i_1\cdots i_n)$, with the amount equal to $1$. Therefore
\begin{equation} 
E_{(i_1)\cdots(i_n)}( \otimes_z |z\rangle^{\otimes c_z}) 
= c_{g(i_1\cdots i_n)} , \label{z2}
\end{equation} 
which, combined with the GRE  subadditivity (\ref{ps}), 
leads to
\begin{equation} 
x_{g(i_1\cdots i_n)} \leq E_{(i_1)\cdots(i_n)}(|\psi\rangle), \label{n}
\end{equation}
which represents $N-1$ inequalities, each of which is
 between the relative
number of copies of a GHZ-like state and a corresponding
entanglement in $|\psi\rangle$. 
The case of $n=2$ is already contained in Eq.~(\ref{ine}). 
So in addition to $M_{bGa}$ equations and 
$M_{bG}-M_{bGa}$ inequalities concerning the bi-GP entanglements, 
we obtain extra $N-2$ inequalities  
concerning entanglement {\em among} more than two
parties, with each party as a GP. 

Combining Eq.~(\ref{n})  with Eq.~(\ref{bga1}), one  thus  obtains 
inequalities for different entanglements in $|\psi\rangle$,
which  can be generally written as 
\begin{equation}
E_{(i_1\cdots i_r)(i_{r+1}\cdots N)} (|\psi\rangle) \leq 
\sum E_{(j_1)\cdots(j_l)(j_{l+1})\cdots(j_k)}(|\psi\rangle), \label{gen}
\end{equation}
for {\em all}  $1 \leq r \leq N-1$, where 
the summation is over all such entanglements for which
$(j_1 \cdots j_l)$ is a subset of GP $(i_1\cdots i_r)$ while 
$(j_{l+1}\cdots j_k)$ is a subset of GP $(i_{r+1}\cdots N)$. 
For example, in the case of $N=3$,  Eq.~(\ref{gen}) represents
\begin{equation}
E_{(12)(3)} (|\psi\rangle) 
\leq E_{(1)(3)}(|\psi\rangle)+E_{(2)(3)}(|\psi\rangle)
+E_{(1)(2)(3)} (|\psi\rangle),
\end{equation}
and the other two inequalities with the party labels cycled. 
{\em Without the appearance of numbers of copies of different states},
inequalities represented by 
Eq.~(\ref{gen})  are entirely the properties of $|\psi\rangle$.

As the primary result of  this section, Eq.~(\ref{gen}) gives
constraints for any state which can be generated from 
${\cal G}(N)$ by reversible LOCC.   

Any violation of the constraints Eq.~(\ref{gen}), or any inconsistency
of any subset of the inequalities of $\{x_z\}$ 
given by Eq.~(\ref{bg1}), Eq.~(\ref{bga1}) and
Eq.~(\ref{n}), would disprove that ${\cal G}(N)$ is a RLEGS.
For this purpose, one could also reduce some 
inequalities concerning  bi-GP entanglements 
not involving all the parties to equalities by
setting these entanglements in $|\psi\rangle$ to zero, and then look
for a counterexample for which the system of equations of $\{x_z\}$
has no solutions, which would signal the 
frustration  in those constraints on the relation between  different
entanglements. However, 
we have not yet found a
counterexample.

\section*{6. Summary}

To summarize,  we sort out all entanglements
in  a  $N$-partite system, where $N \geq 2$.
by associating  an entanglement with each different
{\em set of at least two  GPs}.
The notion of local operation is extended to generalized local
operation  on a GP. For each different set
of GPs, there is a different kind of 
GLOCC. A process of LOCC  is a special process of  GLOCC for any
partition
of these parties into GPs.
The LOCC non-increasing principle implies that
each entanglement involving all the parties is 
invariant under local unitary
transformations  and is non-increasing under LOCC for
all the parties. 
The  so-called 
true $n$-partite entanglement 
is identified as  the  entanglement among all  the $n$ parties,
with each party  as a GP.
A GRE is introduced  as a measure for  each
entanglement.  A GRE is subadditive and monotonic. 
GRE of each entanglement is 
on average  conserved in a reversible LOCC  process.
Then we consider  a combination 
of GHZ-like states states,  with each of which
entering with a certain copies. For this combination, 
additivity holds for 
the GRE of each bi-GP entanglement
and the GRE of true $n$-partite 
entanglements,  $2 \leq n\leq N$.
Consequently,  if a number of copies of a pure state
is generated from the set of all GHZ-like states, ${\cal G}(N)$,
the number of copies of different states concerned have to satisfy 
a series of inequalities.  Moreover, for the entanglement between 
a GP and its complementary GP, the inequalities reduce to equalities. 
Besides, a true $n$-partite entanglement is only contributed 
by the corresponding $n$-GHZ state.
These lead to a series of inequalities between different 
entanglements of  
any state which can be  generated from ${\cal G}(N)$ under reversible LOCC. 
Any  violation of 
these  relations would indicate that 
${\cal G}(N)$ is not a RLEGS.

Finally we mention 
that one can measure each different entanglement
by determining how many copies of one 
{\em generalized}  GHZ-like state~\cite{gghz}
shared by a set of GPs, can be distilled
from how many copies of the concerned multipartite state,
by using the corresponding  GLOCC. 

I thank David DiVincenzo for a very useful comment on 
an earlier version of this work and 
referring to Ref.~\cite{werner}.
I also thank  Gerard Milburn for discussions, and anonymous 
referees
for comments.  This work is partly  an output from project
activity funded by The Cambridge MIT Institute Limited. 

Note added:  It is claimed in \cite{acin} that for three parties,
the set of GHZ and EPR states is not a RLEGS.


\begin{thebibliography}{99}
\bibitem{gghz} A generalized GHZ-like state, shared by GPs
$(1_1 \cdots 1_{n_1})$,
$(2_{1} \cdots 2_{n_2})$, $\cdots$ and $(m_1,\cdots,m_{n_m})$,  is
defined as 
$\frac{1}{ \sqrt{ 2^{n_{min}} } }
\sum_{k=1}^{2^{n_{min}}} \phi_k(1) \cdots \phi_k(m)$,
where  $n_{min}$ is the minimum among $n_q$ ($q=1,\cdots,m$),
$\phi_k(q)$ is a basis state of GP $q$.
\end{thebibliography}
\end{document}